\documentclass[11pt]{article}
\usepackage[margin=1in]{geometry}
\usepackage[utf8]{inputenc}
\usepackage[T1]{fontenc}

\usepackage{wrapfig, braket, float, placeins, colortbl, caption, subcaption, tcolorbox, hhline, pgfgantt, arydshln, amsmath, physics, makecell, amssymb, changepage, siunitx, url, pdfpages, cutwin, graphicx, textcomp, titlesec, lipsum, esvect, pdfpages}

 \DeclareUnicodeCharacter{221A}{$\sqrt{}$}
\DeclareUnicodeCharacter{2202}{$\partial$}

\newcommand{\vect}[1]{\mathbf{#1}}
\DeclareSIUnit\oersted{Oe}
\newcommand{\ipc}{$\mathrm{ions/cm^2}$}

\AtBeginDocument{\RenewCommandCopy\qty\SI}

\usepackage[backend=biber, sortcites = true, sorting=none, autocite=superscript, style=numeric-comp, bibstyle=nature, autopunct=false, maxbibnames=20]{biblatex}

\bibliography{references}

\makeatletter
\renewcommand{\maketitle}{\bgroup\setlength{\parindent}{0pt}
\begin{flushleft}
  \textbf{\textsf{\LARGE{\@title}}}
  
  \@author
\end{flushleft}\egroup
}
\makeatother

\AtBeginDocument{\RenewCommandCopy\qty\SI} 

\title{Inducing a Tunable Skyrmion-Antiskyrmion System through Ion Beam Modification of FeGe Films}

\date{}

\author{%
M. B. Venuti,$^{1}$ Xiyue S. Zhang,$^{2}$ Eric J Lang,$^{3}$ Sadhvikas J. Addamane,$^{4}$ Hanjong Paik,$^{6-8}$ Portia Allen,$^{1}$ Peter Sharma,$^{5}$ David Muller,$^{2,6}$ Khalid Hattar,$^{4,9}$ Tzu-Ming Lu,$^{4}$ Serena Eley$^{1,10}$\\
$^{1}$Department of Physics, Colorado School of Mines, Golden, CO 80401\\
$^{2}$Department of Applied Physics, Cornell University, Ithaca, NY 14853\\
$^{3}$Department of Nuclear Engineering, University of New Mexico, Albuquerque, NM 87131\\
$^{4}$Center for Integrated Nanotechnologies, Sandia National Laboratories, Albuquerque, NM 87123\\
$^{5}$Sandia National Laboratories, Albuquerque, NM 87123\\
$^{6}$Platform for the Accelerated Realization, Analysis, and Discovery of Interface Materials, Cornell University, Ithaca, NY 14853, USA\\
$^{7}$School of Electrical \& Computer Engineering, University of Oklahoma, Norman, OK 73019\\
$^{8}$Center for Quantum Research and Technology, University of Oklahoma, Norman, OK 73019 \\
$^{9}$Department of Nuclear Engineering, University of Tennessee, Knoxville, TN 37996 \\
$^{10}$Department of Electrical \& Computer Engineering, University of Washington, Seattle, WA 98155\\
    \underline{$^{1,10}$serename@uw.edu}\\ }

\titleformat{\section}
  {\Large\sffamily\bfseries}{\thesection}
  {}
  {0pt}
  {}
\titleformat{name=\section,numberless}
  {\Large\sffamily\bfseries}
  {}
  {0pt}
  {}
\titleformat{\subsection}
  {\sffamily\bfseries}{\thesubsection}
 {}
  {}

\titleformat{\subsubsection}
  {\sffamily\small\bfseries\itshape}
  {\thesubsubsection}
  {}
  {}

\titleformat{\paragraph}[runin]
  {\sffamily\small\bfseries}{}{0em}{ }
\begin{document}

\maketitle

\section*{ABSTRACT}

Skyrmions and antiskyrmions are nanoscale swirling textures of magnetic moments formed by chiral interactions between atomic spins in magnetic non-centrosymmetric materials and multilayer films with broken inversion symmetry. These quasiparticles are of interest for use as information carriers in next-generation, low-energy spintronic applications. To develop skyrmion-based memory and logic, we must understand skyrmion-defect interactions with two main goals --- determining how skyrmions navigate intrinsic material defects and determining how to engineer disorder for optimal device operation.  Here, we introduce a tunable means of creating a skyrmion-antiskyrmion system by engineering the disorder landscape in FeGe using ion irradiation. Specifically, we irradiate epitaxial B20-phase FeGe films with 2.8 \unit{\mega \electronvolt} Au$^{4+}$ ions at varying fluences, inducing amorphous regions within the crystalline matrix.  Using low-temperature electrical transport and magnetization measurements, we observe a strong topological Hall effect with a double-peak feature that serves as a signature of skyrmions and antiskyrmions.  These results are a step towards the development of information storage devices that use skyrmions and anitskyrmions as storage bits and our system may serve as a testbed for theoretically predicted phenomena in skyrmion-antiskyrmion crystals.

\section*{Introduction}

Exchange interactions between neighboring spins within ferro- and antiferromagnetic materials lead to a collinear arrangement of lattice spins that is described by the symmetric Hamiltonian. Strong spin-orbit interactions and broken inversion symmetry can give rise to an antisymmetric exchange interaction called the Dzyaloshinskii-Moriya interaction (DMI), which contributes an additional term to the magnetic Hamiltonian, $H_{DMI}= -\vect{D_{ij}}(\vect{S_i} \times \vect{S_j})$, for $D_{ij}$ is the DMI interaction vector, $\vect{S_i}$ and $\vect{S_j}$ represent neighboring spins, and $J$ is the exchange constant. Favoring canting of neighboring spins, the DMI begets a panoply of noncollinear magnetic textures. Of particular interest, it can result in the formation of skyrmions --- various winding configurations of magnetic moments that can have chiral or achiral forms.  Within a material-dependent temperature and magnetic field range, this skyrmion phase is stable, that is, there is an effective energy barrier against the spins within each skyrmion aligning with the ferromagnetic background (topological protection).  B20 phase FeGe, for example, has garnered significant interest because its noncentrosymmetric B20 cubic structure enables Dzyaloshinskii-Moriya interactions that result in skyrmion formation.

Anisotropic DMIs can engender an antiskyrmion -- the antiparticle of a skyrmion -- in which the magnetic moments are reversed. Such localized magnetic textures are categorized based on a topological charge (also known as the skyrmion number), $Q = (1/4\pi)\int{\! \vect{m(r)} \cdot [\partial{_x\vect{m(r)}} \times \partial{_y\vect{m(r)}}] \ \mathrm{d}x \ \mathrm{d}y}$, in which $\vect{m(r)}$ is the magnetization unit-vector field \cite{Zheng2022, Nagaosa2013, Koshibae2016}. Accordingly, antiskyrmions have opposite topological charge from skyrmions, e.g. -1 for skyrmions and +1 for antiskyrmions \cite{Koshibae2016, Hoffmann2017, Zheng2022}.  

Skyrmions and antiskyrmions are of interest for next-generation spintronic devices \cite{Yokouchi_2022,Hoffmann_2021,luo_2021,Song_2020,zhang_2015,tomasello_2014, Zhang_2020,Silva_2020, Zhang2015maglogic,Leonov2017,Jiang2017}. This is because they are predicted to require less energy to manipulate than charge currents in semiconductor or ferromagnetic domain-based logic \cite{Yu_2010, Nagaosa2013, Yu2012, Fert2013, Jonietz2010}, and also have considerable stability due to their topological protection. Implementing skyrmion-based spintronics necessitates understanding how to controllably manipulate these magnetic quasiparticles. It also requires understanding how the skyrmion lattice is affected by disorder \cite{fernandes}, dictating how skyrmions will either maneuver around or be pinned by energy barriers within the material's disorder landscape. This goes beyond considering how skyrmions navigate intrinsic material disorder to include the utility of artificially introduced defects. Precisely-engineered disorder landscapes can be used to control the dynamics of skyrmions in disparate ways; for example, the Magnus force can induce effective repulsion from defects and effectuate an increase in net mobility, while strategic choices of defect species may cause a net attraction, possibly pinning skyrmions \cite{reichhardt_2022}.  Such control could be exploited in device design \cite{arjana_2020}.

Here, we report a systematic investigation of the robustness of the skyrmion phase in FeGe to disorder. To this end, we irradiate epitaxial FeGe films with 2.8 \unit{\mega \electronvolt} Au$^{4+}$ ions, varying the ion fluence to control the densities of induced vacancy clusters. We subsequently capture the disorder-induced evolution of the skyrmion lattice phase in the field-temperature phase diagram using magnetization and Hall effect measurements. We find that the irradiation process induces amorphous regions within a crystalline FeGe matrix and reveal the emergence of a potential skyrmion-antiskyrmion compound system when we pass a certain amorphization threshold.

\section*{Results and Discussion}

\begin{figure}[h!]
    \centering
    \includegraphics[width=0.5\textwidth]{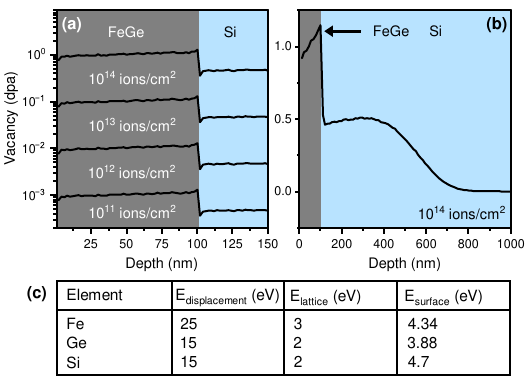}\\ \vspace{.3cm}
    \caption{\textbf{SRIM simulations showing the damage per atom generated by uniform $\mathrm{Au^{4+}}$ ion irradiation at varying fluences.} (a) Displacement profiles for $10^{11}$, $10^{12}$, $10^{13}$, and $10^{14}$ \ipc. (b) Vacancy profile highlighting Bragg peak well within depth of Si substrate. Note that all fluences show a Bragg peak in the same location. (c) Displacement energies ($E_{displacement}$), lattice binding energies ($E_{lattice}$), and surface binding energies ($E_{surface}$) for Fe, Ge, and Si used in the simulations, based on parameters from Ref. \cite{MaterialsProject_FeGe}.} 
    \label{fig:SRIM}
\end{figure}

We sputtered 55 \unit{\nano \meter}-thick epitaxial FeGe films with $\mathrm{P2_13}$ cubic structure (B20 structure) on Si(111) substrates. The growth process is summarized in the Methods section and further detailed elsewhere \cite{PhysRevMaterials.2.074404}. FeGe was ideal for this study because it can be grown epitaxially, therefore with minimal disorder compared to sputtered multilayers, and has been repeatedly shown to host skyrmions associated with a topological Hall effect signature \cite{Leroux2018, gallagher, Ahmed_2018, Budhathoki_2020, Huang_2012} and verified through Lorentz transmission electron microscopy \cite{Yu_2010, Zhao2022, Streubel, Ahmed_2018, Shibata2013} and electron phase microscopy \cite{Kotani_2018}. Moreover, it has a relatively high Curie Temperature $T_C$, measured to be 276.3 K in our pristine (unirradiated) films (see Supplementary Materials Fig. \ref{fig:Curie_Temp}), consistent with the ordering temperature found in other studies of films \cite{PhysRevB.90.024403} and near the bulk value \cite{PhysRevLett.107.127203, BLebech_1989, TEricsson_1981}. Using a 6 MV HVEE EN Tandem Van de Graaff Accelerator at the Sandia Ion Beam Lab, we uniformly irradiated multiple \qtyproduct{6 x 6}{\milli \meter} samples with 2.8 MeV $\mathrm{Au^{4+}}$ ions, each irradiated with a different fluence of $10^{11}$, $10^{12}$, $10^{13}$, and $10^{14}$ \ipc \ to produce damage levels of $10^{-3}$, $10^{-2}$, $10^{-1}$, $1$ displacements per atom (dpa), respectively. Figure \ref{fig:SRIM} shows Stopping and Range of Ions in Matter (SRIM) simulations \cite{Ziegler2010} used to identify the appropriate energies and fluences that would induce our desired range of defect densities.  From Fig. \ref{fig:SRIM}(a), we see that the induced defect density is spread uniformly throughout the depth of the FeGe film and Fig. \ref{fig:SRIM}(b) highlights a Bragg peak then vanishing dpa within the Si substrate. Consequently, we see that the majority of the 2.8 MeV $\mathrm{Au^{4+}}$ ions are not implanted within the FeGe, but rather in the substrate.  Lastly, the table in Fig. \ref{fig:SRIM}(c) summarizes the displacement, lattice binding, and surface binding energies used for each species. Refer to the Methods section for more details regarding the irradiation process.

To further characterize the effects of irradiation, we perform cross-sectional scanning transmission electron microscopy (STEM) imaging,  electron energy loss spectroscopy (EELS), and electron diffraction on FeGe lamellae using a Thermo-Fisher Scientific Spectra 300 STEM Microscope at the Platform for the Accelerated Realization, Analysis, and Discovery of Interface Materials (PARADIM) at Cornell University.  The lamellae were prepared with a Thermo Fisher Helios G4 UX focused ion beam (FIB), also at PARADIM. Refer to the Methods section for more details on the FIB and STEM procedures.

\begin{figure}[htb!]
    \centering
    \includegraphics[width=0.5\textwidth]{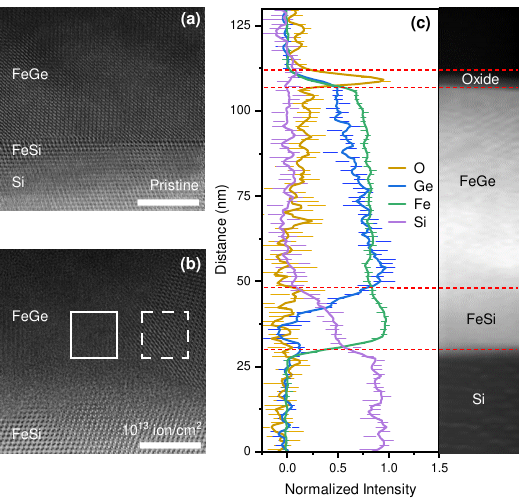}
    \caption{\textbf{Comparison of the microstructure in the irradiated versus the pristine FeGe films.} Cross-sectional STEM image of the (a) pristine FeGe film and (b) film irradiated with $10^{13}$ \ipc. Dashed square encloses an irradiation induced amorphous region and the solid square encompasses an example of a crystalline region.  Scale bars are 5 nm. (c) EELS measurement of the FeGe film irradiated with 10$^{13}$ \ipc. Solid lines show signal intensity for each element; the signal is smoothed using a third-order Savitzky-Golay filter.}
    \label{fig:STEM}
\end{figure}

First considering the pristine (unirradiated) sample, the STEM image in Fig. \ref{fig:STEM}(a) displays an approximately 1 \unit{\nano\meter}-thick FeSi seed layer interlaid between the Si substrate and epitaxial FeGe, as well as clean interfaces. In comparison, Fig. \ref{fig:STEM}(b) is a micrograph of the sample that was irradiated with $10^{13}$ \ipc, revealing crystalline and irradiation-induced amorphized regions. We then collected EELS measurements on the same film, which are presented in Fig. \ref{fig:STEM}(c) and show an approximately 5 nm-thick surface oxide layer, which we observed in all samples, and a homogeneous concentration of Fe and Ge throughout the film thickness. Scans for Au atoms revealed no Au concentrations above measurement resolution, consistent with simulations showing that Au was not implanted within the FeGe.
 
\begin{figure}[ht!]
    \centering
    \includegraphics[width=0.5\textwidth]{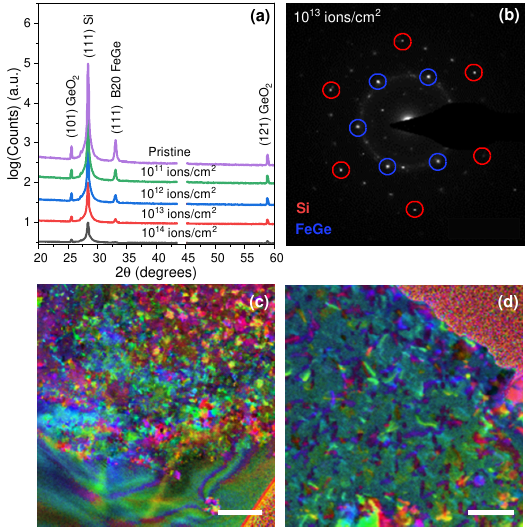}
    \caption{\textbf{Lattice structure mapping and grain identification of pristine and irradiated FeGe films}. (a) Cu-$K_\alpha$ XRD patterns of pristine and irradiated films. Void regions are due to removal of peaks produced by sample mounting hardware. (b) Select Area Electron Diffraction data of film irradiated at $10^{13}$ \ipc \ along the (111) plane. Note the diffuse ring indicative of amorphous material overlaid on the crystalline FeGe peaks. Scale bar is 5 $\mathrm{nm^{-1}}.$ (c) Grain orientation map of film irradiated at $10^{13}$ \ipc. Colored granular regions in top half of image correspond to individual crystalline and amorphous grains. (d) Grain orientation map of film irradiated at $10^{14}$ \ipc. Void region in top right of image is empty space off of the sample surface. Scale bars in (c-d) are 5 \unit{\nano\meter}. }
    \label{fig:diffraction}
\end{figure}

The STEM and EELS results show a surface oxide, and that irradiation induces various sizes and densities of amorphized regions. To identify the lattice structures and map grains, we performed x-ray diffraction (XRD) and scanning nanobeam electron diffraction (NBED) shown in Figs. \ref{fig:diffraction}(a), (b), and (c,d), respectively. The XRD data in Fig. \ref{fig:diffraction}(a) shows the expected (111) Si and (111) B20 FeGe peaks, and identifies the oxide as (101) $\mathrm{GeO_2}$ and (121) $\mathrm{GeO_2}$. The (111) peak appears at 33.1\textdegree{} which corresponds to a measured lattice constant of $4.683 \pm 4.121\times 10^{-4}$ nm. This corresponds to a strain of 0.1\% compared to the bulk lattice constant of 4.679 nm. The full-width-half-maximum of the (111) peak was measured to be $0.233 \pm 0.001$\textdegree{}.  Additionally, observe that as irradiation fluence is increased, there is a gradual decrease in the height of both the crystalline Si and the (111) B20 FeGe peaks. This is suggestive of increased sample amorphization, and in the film irradiated at 10$^{14}$ \ipc, nearly complete amorphization of the B20 FeGe as the peak completely disappears. 

Amorphorization is also evident from the electron diffraction data.  Notably, the NBED data for the sample irradiated at 10$^{13}$ \ipc, Fig. \ref{fig:diffraction}(b), shows Si and FeGe peaks rotated 30\textdegree{} from one another, with a diffuse ring overlapping the FeGe peaks, indicative of partially amorphized FeGe. The colorful mosaics displayed in Figs. \ref{fig:diffraction}(c,d) are false-color grain maps created by Exit Wave Power Cepstrum (EWPC) performed on 4D-STEM NBED data. EWPC cleanly decouples the tilt, thickness, and shape factor information from the lattice structure in thin film diffraction patterns, greatly simplifying the classification of small grain structures at arbitrary orientations. Further details regarding the technique are outlined in Ref. \cite{Zhang2020}. From the grain maps, areas with similar Bragg peaks rotation and periodicity are assigned the same color, thus the color map reveals the distribution and sizes of grains. Regions without clear boundaries are amorphized areas [e.g. the dull green region in Fig. \ref{fig:diffraction}(d)].

\subsection*{Topological Hall Effect Measurements}

To map the magnetic phase diagram, we perform Hall effect measurements at variable temperatures and magnetic fields. We then use the topological Hall effect (THE) as the hallmark of the emergence of nontrivial spin textures \cite{WANG2022100971} \--- namely, skyrmions. Noncollinear spin textures engender an effective magnetic field that deflects charge carrier motion, resulting in a THE. Such textures include topological states (different types of skyrmions, antiskyrmions), spin chirality in Kagome and triangular lattices, and right- or left-handed chiral domain walls \cite{WANG2022100971}. In FeGe, the THE is a recognized signature of skyrmions \cite{Leroux2018, Ahmed_2018, Huang_2012, gallagher, Budhathoki_2020}.

To this end, we conduct electrical transport measurements on microfabricated six-point Hall bar devices and out-of-plane magnetization $M$ measurements on approximately \qtyproduct{3x3}{\milli\meter} unpatterned films. We directly measure the total Hall resistivity $\rho_{xy}$, which consists of contributions from three different effects: the ordinary Hall effect, the anomalous Hall effect, and the topological Hall effect. It can therefore be expressed as the sum of three terms \cite{Huang_2012, Surgers2014, WANG2022100971, Nagaosa_2012, PhysRevB.90.024403}

\begin{equation}\label{eq:rho_THE}
    \rho_{xy} =  \rho_{OHE} +  \rho_{AHE} + \rho_{THE},
\end{equation}

for ordinary Hall resistivity $\rho_{OHE} = R_0 \mu_0H$, ordinary Hall coefficient $R_0$, externally applied out-of-plane magnetic field $\mu_0H$, anomalous Hall resistivity $\rho_{AHE}=R_{AH} M(T,H)$, anomalous Hall coefficient $R_{AH}$, and topological Hall resistivity $\rho_{THE}$. Through the anomalous Hall effect contribution $R_{AH}(\rho_{xx})$, $\rho_{xy}$ also depends on the longitudinal resistivity $\rho_{xx}$, which we measure in conjunction with $\rho_{xy}$, and the measured sample out-of-plane magnetization $M(T,H)$. 

To extract $\rho_{THE}$, we calculate and subtract both $\rho_{OHE}$ and $\rho_{AHE}$ from $\rho_{xy}$. First, recognizing that in saturation, all topological magnetic textures vanish and $\rho_{THE}$ is zero, we find that

\begin{equation}\label{eq:rho_THE2}
    \frac{\rho_{xy}}{\mu_0H} = R_0 + R_{AH} \frac{M}{\mu_0H},\ H > H_{sat}.
\end{equation}

\noindent After antisymmetrizing the raw resistivity and magnetization data, we determine $R_0$ and $R_{AH}$ by performing a linear fit to $\rho_{xy}/H$ versus $M/\mu_0H$, noting $R_0$ as the intercept and $R_{AH}$ as the slope, as seen in Equation \ref{eq:rho_THE2}. From the extracted coefficients, we can then calculate $\rho_{OHE}$, $\rho_{AHE}$, and subsequently $\rho_{THE}$. This topological Hall resistivity is proportional to the net topological charge, such that larger $|\rho_{THE}|$ corresponds to a higher density of topological magnetic textures.

\begin{figure}[htb!]
    \centering
    \includegraphics[width=0.5\textwidth]{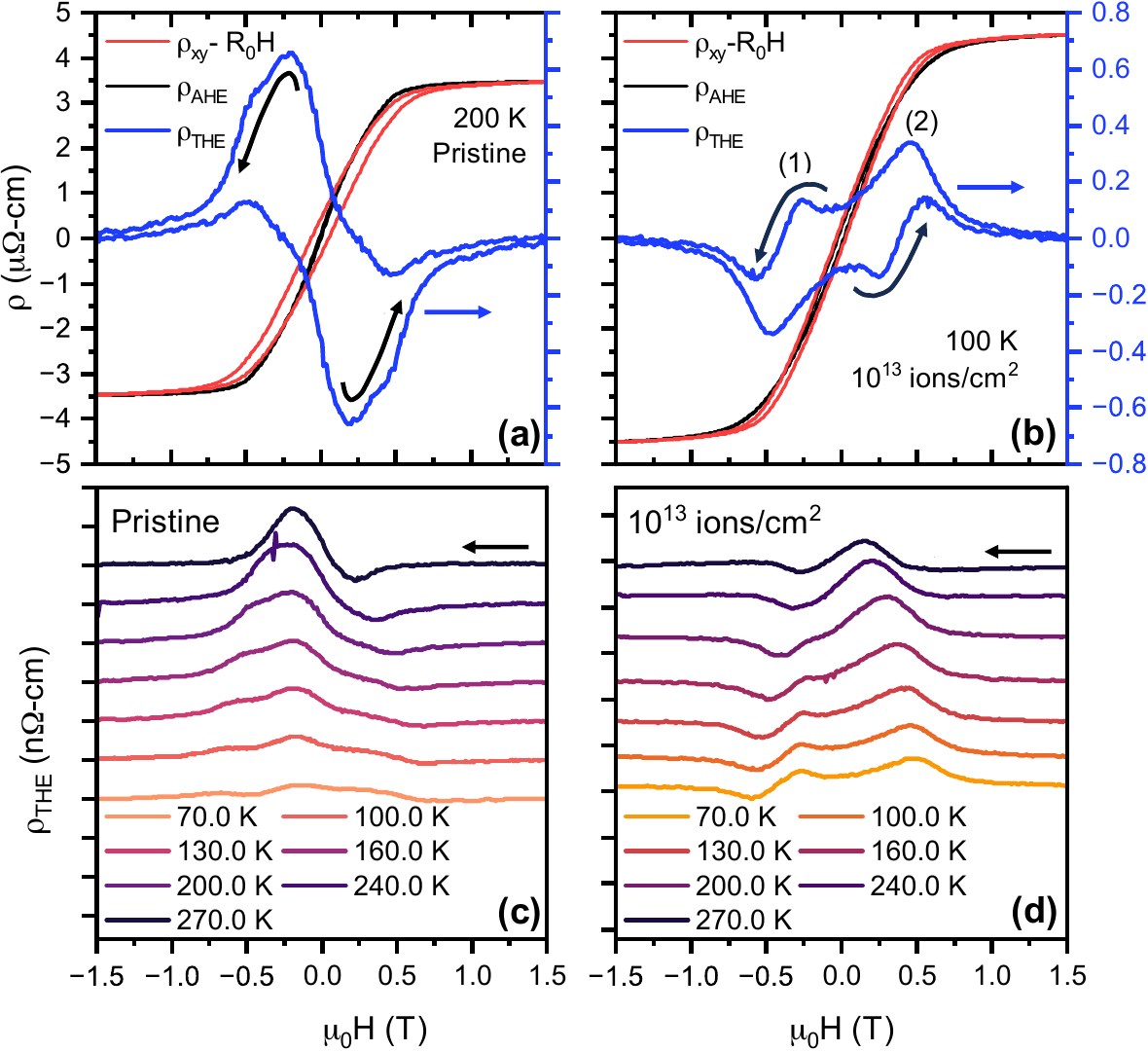}
    \caption{\textbf{Topological Hall effect in pristine and irradiated FeGe films.} (a-b) Subtracting OHE and AHE components to extract $\rho_{THE}$. Blue $\rho_{THE}$ are plotted versus the right axes in both plots. (c-d) Isothermal plots of $\rho_{THE}$ versus applied magnetic field for a (c) pristine and a FeGe film that was irradiated at $\mathrm{10^{13}}$ \ipc. The field is swept from high to low for all curves. Arbitrary offsets added for clarity. Tick marks on y-axis have spacings of 500 \unit{\nano\Omega.cm}.}
    \label{fig:Composite RhoTHE}
\end{figure}

Figure \ref{fig:Composite RhoTHE}(a) shows the field-dependent Hall resistivities for the pristine FeGe film at 200 \unit{\kelvin}. The blue curve is the extracted $\rho_{THE}$, which shows one prominent positive-valued peak in the negative-field region as the applied magnetic field is swept from high to low. This is the typical shape expected from FeGe \cite{gallagher, Ahmed_2018, Budhathoki_2020, Huang_2012}, and we attribute this peak to the presence of skyrmions.  The evolution of the peak with temperature is shown in Figs. \ref{fig:Composite RhoTHE}(c) and \ref{fig:THE Combined All}(b); the THE peak persists down to the lowest measurement temperature of 10 \unit{\kelvin}. Similar results are seen for the film irradiated at 10$^{11}$ \ipc \ and 10$^{12}$ \ipc, as shown in Figs. \ref{fig:THE Combined All}(c) and \ref{fig:THE Combined All}(d).

As defect density increases, we see the emergence of a second peak \--- in the positive-field region \--- as seen in Fig. \ref{fig:Composite RhoTHE}(b). Specifically, we see the single-peak behavior in the unirradiated film and samples irradiated with 2.8 MeV Au$^{4+}$ fluences below $10^{13}$ \ipc, corresponding to damage levels below $\mathrm{10^{-1}}$ dpa, and double-peak features at and above $10^{13}$ \ipc.  We label the peaks (1) and (2) in Fig. \ref{fig:Composite RhoTHE}(b), and find that the data has a similar shape to that observed in the inverse Heusler system Mn$_{1.4}$Pt$_{0.9}$Pd$_{0.1}$Sn \cite{sivakumar} and that peak (1) is consistent with the presence of skyrmions.  To identify the origin of peak (2), we consider the relative properties of another magnetic texture that may be present.

The topological Hall resistivity is directly proportional to the topological charge density, which can be written as a product of polarization $p$ and vorticity $W$ \cite{hrkac_2015},

    \begin{equation}\label{eq:pW}
        \rho_{THE} = R_{THE}\langle n \rangle =pW,
    \end{equation}
  
where $R_{THE}$ is the topological Hall coefficient and $\langle n \rangle$ is the total topological charge density of the sample.  Note that $R_{THE}$ is proportional to both the carrier density and the carrier polarization. Moreover, the carrier density is invariant under field sweeps, such that any change in the sign of $\rho_{THE}$ during a field sweep is due solely to a change in the sign of $\langle n \rangle$. 

\begin{figure}[htb!]
\centering
\includegraphics[width=\textwidth]{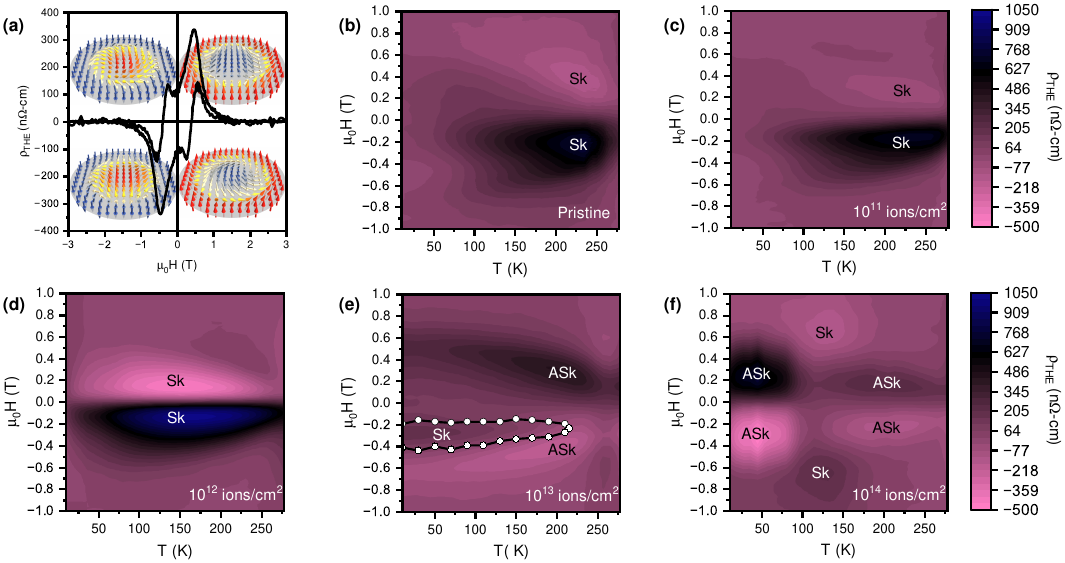}
\caption{\textbf{Magnetic phase diagram in pristine and irradiated FeGe films.}(a) Magnetic field-dependent topological Hall resistivity of FeGe film irradiated at $\mathrm{10^{13}}$ \ipc at 100 K. Each quadrant is labelled with the corresponding topological spin texture type and polarization associated with the peak or dip. (b-f) Two-dimensional heat maps of $\rho_{THE}$ (color scale) for each sample. The phase boundary in (e) was found by plotting inflection points in $\rho_{THE}$. All diagrams are plotted with field sweeping from high to low.} 
\label{fig:heatmaps}
\end{figure}

The magnetic textures associated with each peak must have topological charges $\langle n \rangle$ of the same sign, yet polarities $p=\pm 1$ of opposite sign. This is because, first, the $\rho_{THE}$ of the dual peaks have the same sign, hence $\langle n \rangle$ has the same sign for each peak. Second, the magnetization $M$ for each peak has opposite sign, such that the sign of the polarity must be opposite. As we see from Eq. \ref{eq:pW}, given that the two textures both produce positive $\rho_{THE}$, yet polarities of opposite sign, then they must have vorticities $W=\pm 1$ of opposite sign. This reasoning \cite{sivakumar} therefore suggests that as peak (1) is representative of skyrmions (with vorticity +1), peak (2) may be caused by a texture of opposite polarity and vorticity \-1 \--- antiskyrmions. Fig. \ref{fig:heatmaps}(a)
is a cartoon depicting the type of topological spin texture responsible for each peak and dip in $\rho_{THE}(H)$, showing skyrmions of polarity +1 (upper left quadrant), antiskyrmions of polarity -1 (upper right quadrant), antiskyrmions of polarity +1 (lower right quadrant), and skyrmions of polarity -1 (lower left quadrant). 

Antiskyrmions have been observed in ultrathin multilayer films with interfacial DMIs \cite{Zhang2016, PhysRevB.95.214422}, Heusler compounds with D$_{2d}$ symmetry \cite{Nayak2017, Peng2020, Shimizu2022, Saha_2019, Jamaluddin2019, Jena2020, Jena2020b, PhysRevB.101.014424, PhysRevB.102.174402, Ma2020, PhysRevB.103.184411}, and a schreibersite with $S_4$ symmetry \cite{Karube2021}. They have also been predicted to form in dipolar magnets \cite{Koshibae2016}, chiral magnets in tilted magnetic fields \cite{PhysRevB.101.064408}, and ultra-thin magnetic films grown on semiconductors or heavy metals with $C_{2v}$ symmetry \cite{Hoffmann2017}. One study predicted that antiskyrmions would be unstable \--- immediately vanishing --- in chiral magnets \cite{Koshibae2016}, however they have in fact recently been observed via Lorentz Transmission Electron Microscopy (LTEM) in a 70 nm-thick lamella of B20-type FeGe \cite{Zheng2022} and in amorphous $\mathrm{Fe_{0.52}Ge_{0.48}}$, with antiskyrmion stabilization occurring as a result of randomized DMI direction from amorphization \cite{Streubel}. 

From examining $\rho_{THE}$ plotted as a function of both temperature and field, we can construct the effective phase diagrams that are shown in Fig. \ref{fig:heatmaps}(b-f).   Based on the peak positions in the topological Hall resistivity, we label the corresponding magnetic textures -- Sk (skyrmion) and ASk (antiskyrmion). Accordingly, we see evidence of skyrmions alone in the pristine sample (crystalline FeGe) and film irradiated at $10^{11}$ \ipc in Figs. \ref{fig:heatmaps}(b) and (c), respectively.  In the film irradiated at $10^{12}$ \ipc \ [results shown in Figs. \ref{fig:heatmaps}(b)], the skyrmion phase broadens out to lower temperatures, suggesting that skyrmions are forming at these lower temperatures in the amorphized regions. Since composition of amorphous material increases with increased fluence, we deduce that the skyrmion phases in both amorphous FeGe and crystalline FeGe are overlapping to produce the appearance of one broad skyrmion phase. If we compare negative-field skyrmion peak heights between the different samples, we observe that the peak height for the pristine sample and that of the sample irradiated at $1\times 10^{11}$ \ipc{} are similar, and a 1.8$\times$ increase betweenthe peak height for the sample irradiated at $1\times 10^{11}$ and that irradiated at $1\times 10^{12}$ \ipc{}. Note that the peak height is directly proportional to skyrmion density.
The marked decrease in skyrmion density in Figure \ref{fig:heatmaps}(e) is predicted to be from a lack of skyrmion-favoring crystalline FeGe.

As irradiation fluence increases to $10^{13}$ \ipc (increasing the density of amorphous grains) we see the emergence of the double-peak feature. Figure \ref{fig:heatmaps}e displays the corresponding broad antiskyrmion phase down to single Kelvin temperatures, and the skyrmion phase emerges at temperatures below 210 \unit{\kelvin} and persists down to temperatures lower than in the pristine film.  We propose that the magnetic ordering temperature of the skyrmion phase is less than in the pristine film, in which it orders at $\sim$280 \unit{\kelvin}, due to a decrease in uniaxial anisotropy resulting from 2.8 MeV Au$^{4+}$ irradiation.  This implies that the crystalline regions host skyrmions whereas the amorphous regions hosts both skyrmions and antiskyrmions.  Examining an idealized Hamiltonian \cite{PhysRevApplied.21.034009} of a ferromagnet with DMI vector $\vect{D_{ij}}$, symmetric exchange coefficient $J_{ij}$, and uniaxial anisotropy constant $K$,

\begin{equation}\label{eq:H}
    H_{ij} = -\sum_{<i,j>}J_{ij}(\vect{S}_i\cdot\vect{S}_j) - \sum_{<i,j>}\vect{D_{ij}}\cdot(\vect{S}_i\times\vect{S}_j) -K\sum_{i}(S_i^z)^2-\sum_i \mu_0\mathbf{B}\vect{S}_i^z,
\end{equation}

we expect the exchange coefficient and the DMI vector magnitude $D_{ij}$ to change minimally as a result of amorphization, because they depend on local bond geometries \cite{Dzyaloshinsky_1958,Noodleman_1981}. On the other hand, the uniaxial anisotropy constant $K$ and the direction of the DMI vector $\hat{D_{ij}}$ could change substantially since they depend on long-range order of the lattice, which is broken during amorphization Note that, in equation \ref{eq:H}, $\vect{B}$ is the z-oriented magnetic field, $\vect{S_i^z}$ is the z component of $\vect{S_i}$, and $\mu$ is the magnetic moment of the magnetic atom.

Lastly, as fluence is further increased to $10^{14}$ \ipc, we see predominately antiskyrmions with a narrow skyrmion phase existing between approximately 70 \unit{\kelvin} and 125 \unit{\kelvin} in Fig. \ref{fig:heatmaps}(f). This sample is near-fully amorphized, and the large fluence could have displaced the atoms to the point that short-range-order quantities such as $J$ and $D$ have changed, so magnetic phase boundaries have fully shifted with respect to the lower-fluence samples.

\begin{figure}[htb!]
    \centering
    \includegraphics[width=\textwidth]{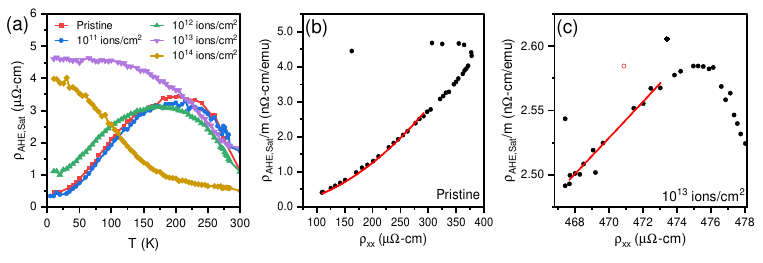}
    \caption{\textbf{Anomalous Hall resistivity in pristine and irradiated FeGe films.} (a) Temperature-dependent $\rho_{AHE}$ in the positive-field saturation region for all FeGe samples.
    (b,c) Fit of Eq. \ref{eq:AHE} to anomalous Hall resistivity divided by magnetic moment vs longitudinal resistivity, for (b) the pristine sample for temperatures less than 130 K and (c) the sample irradiated at $10^{13}$ \ipc~ for temperatures less than 95 \unit{\kelvin}. The open red circle was associated with noise and excluded from from the fit.}
    \label{fig:Msat}
\end{figure}

 Figure \ref{fig:Msat}(a) shows the saturated value of the AHE resistivity, $\rho_{AHE,Sat}$, for the pristine and irradiated samples as a function of temperature. Note that the shape of the curves for the films irradiated at $10^{12}$ \ipc, $10^{13}$ \ipc, and $10^{14}$ \ipc are similar to that found in textured FeGe films \cite{PhysRevB.90.024403}. If we compare $\rho_{AHE,Sat}$ for each sample, we can attribute differences to changes in $K$ and in the direction of $\vect{D_{ij}}$, which we will denote $\hat{D}$. We see that at approximately 210 K, saturation magnetization is effectively identical for all samples containing crystalline FeGe. Neglecting the differences above this temperature, we assume that changes in $K$ and $\hat{D}$ are no longer visible in saturation. Because we have evidence that antiskyrmions may persist at temperatures above 210 K in the sample irradiated at $10^{13}$ \ipc, we can not deduce significant changes in $\vect{D_{ij}}$ in the amorphous regions, and attribute all changes in $\rho_{AHE,Sat}$ to $K$. 

Note that $\rho_{AHE,Sat}$ for the fully amorphized sample is lower and has a much smaller rate of change with respect to temperature. We attribute the behavior of $\rho_{AHE,Sat}$ at temperatures above 210 K to be primarily due to crystalline FeGe in the lower fluence samples, since the curves have a significant negative slope and larger values. The vanishing skyrmion phase above 210 \unit{\kelvin} in the sample irradiated at $10^{13}$ \ipc \ indicates that $K$ is important in skyrmion stabilization in the crystalline phase. The lowering of $K$ in crystalline FeGe is likely due to the randomization of grain direction due to irradiation, and provides an explanation to why the helical ordering temperature is lower in irradiated films.

We now consider how different contributions to the anomalous Hall effect are affected by disorder. The anomalous Hall effect results from the co-action of skew scattering, side-jump scattering, and an intrinsic scattering-independent mechanism (related to the Berry curvature). Accordingly, the anomalous Hall coefficient can be expressed as \cite{Budhathoki_2020, RevModPhys.82.1539, PhysRevB.90.024403}:
% \begin{linenomath}
\begin{equation}\label{eq:AHE}
  % R_{AH} =\alpha \rho_{xx}^2+\beta \rho_{xx}.
  R_{AH} =\alpha \rho_{xx} + (\beta + b)\rho_{xx}^2.
\end{equation}
% \end{linenomath}
\noindent Here, the side-jump scattering (coefficient $\beta$) and intrinsic (coefficient $b$) contributions both depend quadratically on the longitudinal resistivity $\rho_{xx}$, whereas the skew-scattering mechanism (coefficient $\alpha$) depends linearly on $\rho_{xx}$. Side-jump scattering occurs when wave packets formed from spin-orbit coupled Bloch states are scattered by a disorder potential \cite{RevModPhys.82.1539} whereas skew-scattering is an antisymmetric scattering process caused by the effective spin-orbit coupling of the conduction electron or impurity from which it scatters \cite{RevModPhys.82.1539}. 

Previous studies of FeGe films have found that the scattering-independent (intrinsic) mechanism prevails and that contributions from skew-scattering, typically only significant in crystals \cite{RevModPhys.82.1539, PhysRevB.77.165103}, can be neglected \cite{Budhathoki_2020, Tsymbal2011, li_2013, Huang_2012}. This is partially because finite-temperature effects tend to suppress the skew-scattering (linear) contribution \cite{PhysRevLett.103.087206, RevModPhys.82.1539}. Ref. \cite{PhysRevB.90.024403} also found evidence of a suppressed linear contribution in textured FeGe films and that the predominating quadratic dependence of $R_{AH}$ on $\rho_{xx}^2$  was consistent with both the intrinsic and the side-jump mechanisms.

To investigate how the relative contributions of the anomalous Hall effect may change in our FeGe films owing to systematically tuned disorder, we compare $\rho_{AHE, Sat}/m$ versus $\rho_{xx}$ in all films. Figure \ref{fig:Msat}(b,c) displays the results for the pristine film and for the sample irradiated with $10^{13}$ \ipc{}, respectively, whereas the data for the other samples is shown in Supplemental Materials Fig. \ref{fig:Additional_AHE}. In all samples, we see evidence that the relative magnitude of the various contributions to $\rho_{AHE}$  --- skew scattering versus the combined effects of side-jump scattering and intrinsic effects --- are temperature dependent. We therefore restrict our analysis to low-temperature regions where simple second order polynomial fits can be made with a convergence of chi-squared $\chi^2<1\times 10^{-9}$. In the pristine sample, the skew-scattering component $\alpha$ appears to be negligible at temperatures less than $\sim$130 \unit{\kelvin}, as shown in Fig. \ref{fig:Msat}(b). A fit to Eq. \ref{eq:AHE} is performed, omitting the linear term by setting $\alpha \equiv 0$, and noting that a good fit can still be made in its absence. We find that $\beta + b = 2.12\ \mathrm{\mu\Omega^{-1}\mu A^{-1}}$. Similarly, for the sample irradiated with a low dose of $10^{11}$ \ipc, we find negligible $\alpha$. 
However, fits for the samples irradiated with higher dosages yielded non-negligible $\alpha$. For example, for the sample irradiated at $10^{13}$ \ipc{}, we find that $\alpha = -2.19 \times 10^{-3}\mathrm{MA^{-1}nm}$ 
    and $\beta + b = 1.24\ \mathrm{\mu\Omega^{-1}\mu A^{-1}}$.  We thus conclude that irradiation increases the contribution of skew-scattering in our FeGe films.

\section*{Conclusion}
In summary, we induce amorphous regions in epitaxially grown FeGe through 2.8 MeV Au$^{4+}$ ion irradiation, and perform subsequent topological Hall effect measurements to investigate topological spin texture formation. Our measurements show evidence that composite skyrmion-antiskyrmion systems form in the partially-amorphized films, with skyrmions existing in the crystalline phase, and both skyrmions and antiskyrmions in the amorphous phase. Fundamentally, such systems are of interest as testbeds for theoretically predicted phenomena such as spin wave emission by current induced skyrmion-antiskyrmion pair annihilation \cite{Zhang2017b}, a skyrmion-antiskyrmion liquid \cite{PhysRevB.93.064430}, as well as skyrmion-antiskyrmion crystals, interactions, and dynamics \cite{Leonov2015, Hu2017,Ritzmann2018}.

Systems hosting coexisting skyrmions and antiskyrmions are of interest for architectures in which the two particles are used for binary data encoding, for example in skyrmion-antiskyrmion racetrack memory \cite{Hoffmann_2021, Silva_2020} and magnetic logic gates \cite{Zhang2015maglogic}.  Information could be stored based on the topological charge. Given opposite topological charges, skyrmions and antiskyrmions have opposite topological Magnus forces, which cause them to accumulate on opposite sides of the sample (skyrmion Hall effect) and can be exploited for logic operations \cite{Leonov2017,Jiang2017}. Such a system would prove beneficial for spintronic applications requiring the transport of magnetic textures in the absence of transverse Hall motion \cite{Huang_2017}.  The presence of both textures in the amorphous phase also presents an opportunity for simpler fabrication of these devices, since the growth of a pure amorphous phase would not require specific substrates and would be less dependent on the requirement of pristine substrate surface preparation and condition. Amorphous FeGe phases have also been shown to exhibit larger Hall effects thus making them more sensitive to voltage manipulation \cite{Streubel}.  One may also find novel uses for creating precisely-formed amorphous shapes within a surrounding crystalline matrix, which would easily be accomplished by performing irradiations through a hard mask so that only particular sections of films are exposed to incident ions.

\vspace{0.5cm}
\noindent\makebox[\linewidth]{\rule{6.5in}{0.4pt}}

\vspace{-0.5cm}
\section*{Methods}\label{sec:Methods}
\subsection*{Film growth}
The FeGe films measured in this study were grown epitaxially on Si (111) substrates using molecular-beam epitaxy (MBE) in a Veeco GEN10 chamber at the Platform for the Accelerated Realization, Analysis, and Discovery of Interface Materials (PARADIM) at Cornell University. First, a FeSi seed layer is established by depositing a monolayer of Fe onto a $7 \times 7$ reconstructed Si (111) surface then flash annealing at \ang{500} C. Subsequently, B20 FeGe is grown atop the seed layer by codeposition of Fe and Ge sources at \ang{200} C, at 0.5 \AA/s, and in a base pressure of 
\qtyproduct{2e-9}{torr}.  The sources were 40 cc effusion cells that produce relatively uniform films covering an area of 1.5 inches. For more details on a similar process used to grow epitaxial Mn$_x$Fe$_{1-x}$Ge films, see Ref. \cite{PhysRevMaterials.2.074404}. 

\subsection*{Ion Beam Modification}

Ion beam modification was carried out using a 6 MV HVEE EN Tandem Van de Graaff Accelerator at the Sandia National Laboratories Ion Beam Lab. FeGe specimens were sliced into \qtyproduct{0.5 x 0.5}{cm} pieces, adhered to a Si backing plate with double sided carbon tape, inserted into the tandem end station, and pumped to a base pressure of at least \qtyproduct{1e-6}{torr}. Ion irradiation was performed using 2.8 MeV $\mathrm{Au^{4+}}$
 ions with an ion beam current of \qtyproduct{100}{nA} 
 (ion flux of \qtyproduct{1.56e11}{\mathrm{ions/cm^2}} measured at the sample before and after irradiation of the samples. Estimations of the ion beam damage were performed with Stopping Range of Ion in Matter (SRIM) simulations \cite{Ziegler2010}. Based on parameters from Ref. \cite{MaterialsProject_FeGe}, the SRIM simulation used an FeGe density of 8.14 \unit{g/cm^3}, Si density of 2.312 \unit{g/cm^3}, as well as the displacement energies, lattice energies, and surface energies indicated in the table in Fig. \ref{fig:SRIM}(c) in the main text.

\subsection*{Focused Ion Beam} 

We used a cross-sectional focused ion beam (FIB) process to prepare the STEM samples studied for crystal structure characterization and chemical composition analysis, for which the results are displayed in Fig. \ref{fig:STEM}.  The samples used for grain mapping were prepared using a plan-view FIB process.  In both cases, first, we deposit a stack of 20--30 \unit{\nano\meter} of carbon and 0.8--1 \unit{\micro\meter} of Pt on the sample surface to form a protective overlayer. Next, we prepare a lamella using a Ga ion beam. We then adhere it to a needle with a sputtered Pt paste in order to transfer it to a TEM grid. The cross-section or plan-view lamellae are then further milled iteratively on both sides, using a 30 \unit{keV} Ga ion beam, down to a thickness of 200--500 \unit{\nano\meter}, then at 5 \unit{keV} until the thickness of the protective Pt layer becomes less than 20 \unit{\nano\meter}. During the plan-view FIB thinning process, the film surface and substrate are milled at a 2-3$^{\circ}$ angle with respect to the ab-plane to create a smooth gradient of different regions, displayed from the bottom (substrate) to top (film), respectively, in Fig. \ref{fig:diffraction}(c).

\subsection*{Hall Bar Fabrication}
We fabricated 55 \unit{\nano\meter}-thick FeGe films into Hall bars using standard photolithography (laser writer) and broad-beam Ar$^+$ ion milling techniques at the Center for Integrated Nanotechnologies (CINT) at Sandia National Laboratories. 100 \unit{\nano\meter}-thick Au contacts were deposited using e-beam evaporation with a 5 \unit{\nano\meter} Ti sticking layer, followed by lift-off in acetone. 

\subsection*{Magnetometry Measurements}
Magnetization studies were performed using a Quantum Design MPMS3 SQUID magnetometer at the Colorado School of Mines and University of Washington. Samples were cut to approximately \qtyproduct{3.5x3.5}{\milli\meter} to friction-fit within a drinking straw for low-background sample mounting. Hence, in calculations requiring a volume $V$, we used $V \approx 6.7375 \times 10^{-4}$  \unit{mm^3}, considering a nominal film thickness of 55 \unit{\nano\meter}. For all measurements, the magnetic field was applied perpendicular to the film plane and swept in a full hysteresis loop between $\pm$ 3\unit{\tesla}. Within $\pm$ 1.5 \unit{\tesla} we measure using 100 Oe intervals, and 500 Oe intervals beyond  this range. Background from the Si substrate was removed by subtracting a linear fit to the saturated regions of the $M(H)$ curves at each temperature. 

\subsection*{Electrical Transport Characterization}
Electrical transport measurements were performed using three SR830 lock-in amplifiers and two SR560 voltage preamplifiers per Hall bar. Devices were measured at either 37 or 41 \unit{\hertz} with an applied current of $
\sim$3.6 \unit{\micro\ampere} rms. The longitudinal resistivity $\rho_{xx}$ and transverse resistivity $\rho_{xy}$ are measured simultaneously using the lock-in amplifier (and preamplifier), and voltage drop across a 100 $\Omega$ series resistor is recorded using another SR830, ultimately to track the applied current. Voltage preamplifiers were set to a constant gain of 100 and a low-pass filter with a cutoff frequency of 10 \unit{\kilo\hertz}. A time constant of 1 \unit{\second} was used for lock-in measurements. A full wiring diagram is given in the Supplemental Materials as Fig. \ref{fig:HallBarDiagram}.

\section*{Acknowledgements}
This material is based upon work supported by the National Science Foundation under grants DMR-1905909 and DMR-2330562.
This work was performed, in part, at the Center for Integrated Nanotechnologies, an Office of Science User Facility operated for the U.S. Department of Energy (DOE) Office of Science. 
This work also used synthesis and microscopy facilities at the Platform for the Accelerated Realization, Analysis, and Discovery of Interface Materials (PARADIM), which are supported by the National Science Foundation under Cooperative Agreement No. DMR-2039380.
M.B.V. thanks David Lidsky (Sandia National Labs) for assistance in using cleanroom instrumentation at the Center for Integrated Nanotechnologies.

\section*{Data availability}
Data that support the findings of this study are available from the corresponding authors upon reasonable request.

\section*{Author contributions statement}
S.E. conceived the experiments, conducted measurements, and advised on device fabrication, measurements, and data analysis;  M.B.V conducted magnetization measurements, Hall effect measurements, as well as SRIM simulations, and performed the Hall bar microfabrication; X.Z and D.M. imaged the films using transmission electron microscopy measurements and performed the accompanying analysis; E.L. and K.H. irradiated the films; E.L. and S. A. J. performed x-ray diffraction and reflectivity measurements; T.M.L. assisted with microfabrication, electrical transport measurements, and data analysis; P.S. assisted with electrical transport measurements; H. P. grew the FeGe films; P. A. assisted in writing Python scripts for electrical transport measurement.   S.E. and M.B.V. wrote the manuscript. All authors commented on the manuscript. 

\section*{Ethics declarations}
The authors declare no competing interests.

\printbibliography

\pagebreak

\title{Supplemental Materials: Inducing a Tunable Skyrmion-Antiskyrmion System through Ion Beam Modification of FeGe Films}
\maketitle

Here we present additional materials in support of the full publication. Additional electron energy-loss spectroscopy (EELS) measurements were performed on pristine FeGe cross-sections, shown in Figure \ref{fig:pristine_eels}, showing the depthwise FeGe-FeSi-Si films structure. We also provide a combined topological Hall resistivity summary of all samples in the study in the form of full hysteresis loops at varying temperatures, shown in Figure \ref{fig:THE Combined All}. We also provide a circuit diagram of the full experimental setup in collecting the transport data, shown in Figure \ref{fig:HallBarDiagram}. Here, two Stanford Research System (SRS) 830 lock-in amplifiers were used to measure the longitudinal and transverse voltage drops across each Hall bar. Both signals were amplified using SRS 560 voltage pre-amplifiers. A third lock-in amplifier was used to measure the voltage drop across 100 \unit{\Omega} resistor placed in series with the Hall bar to for a precise measurement of the current through the device.

\section*{Curie Temperature}
\setcounter{figure}{0}
\makeatletter 
\renewcommand{\thefigure}{S\@arabic\c@figure}
\makeatother

\begin{figure}[H]
    \centering
    \includegraphics[width=0.5\textwidth]{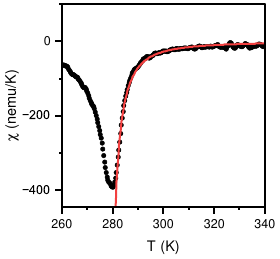}
    \caption{Fit of moment vs temperature to the Curie-Weiss Law $\chi = C(T-T_C)^{-\gamma}$ with a critical exponent $\gamma$ of 1.73 for the pristine sample. The extracted Curie temperature was found to be 276.3 K.}
    \label{fig:Curie_Temp}
\end{figure}

\section*{Electron Energy Loss Spectroscopy (EELS)}

\begin{figure}[H]
    \centering
    \includegraphics[height=0.5\textwidth]{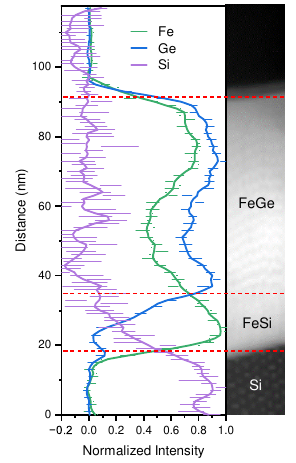}
    \caption{Atomic compositions as a function of depth for a pristine FeGe cross-section.}
    \label{fig:pristine_eels}
\end{figure}

\section*{Topological Hall Measurements}
\begin{figure}[H]
    \centering
    \includegraphics[width=\textwidth]{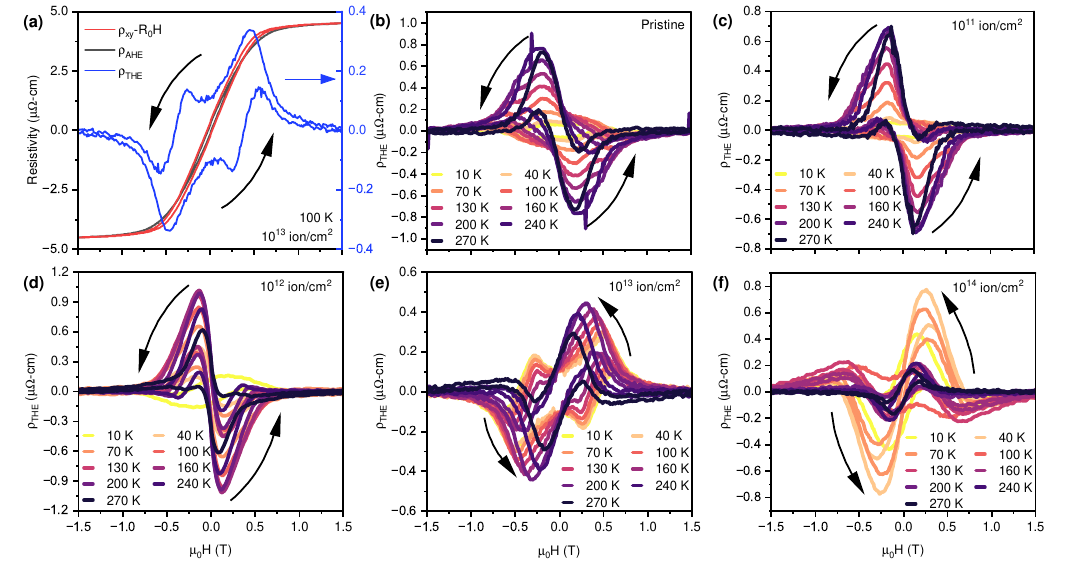}
    \caption{Composite $\rho_{THE}$ for all samples. We see the emergence of the double-peak feature in (e) and (f).}
    \label{fig:THE Combined All}
\end{figure}

% \section*{Magnetization}
% \textcolor{red}{Make big figure with a-e for $\rho$ vs H at different T as one figure. Make a figure with parts a and b with $\rho_{sat} vs T$ and zoomed $\rho_{AHE} vs H$ as another figure. We can talk about the 2-part figure in more detail. }

\section*{Additional Anomalous Hall Resistivity Fits}
\begin{figure}[H]
    \centering
    \includegraphics[width=\textwidth]{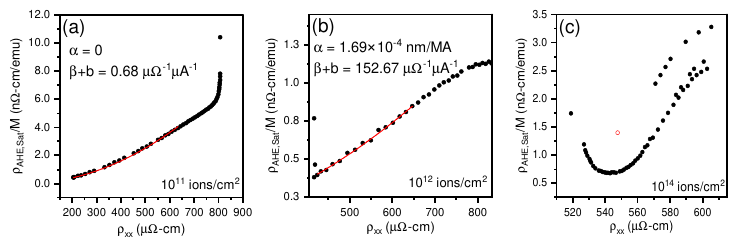}
    \caption{Fit of Eq. \ref{eq:AHE} to anomalous Hall resistivity divided by magnetic moment vs longitudinal resistivity,
for the (a) $10^{11}$, (b) $10^{12}$, and (c) $10^{14}$ \ipc{} samples. No fit is performed in (c) because the low-resistivity data does not fit the model described in Eq \ref{eq:AHE}}
    \label{fig:Additional_AHE}
\end{figure}

\section*{Electrical Transport Measurements}
\begin{figure}[H]
    \centering
    \includegraphics[width=\textwidth]{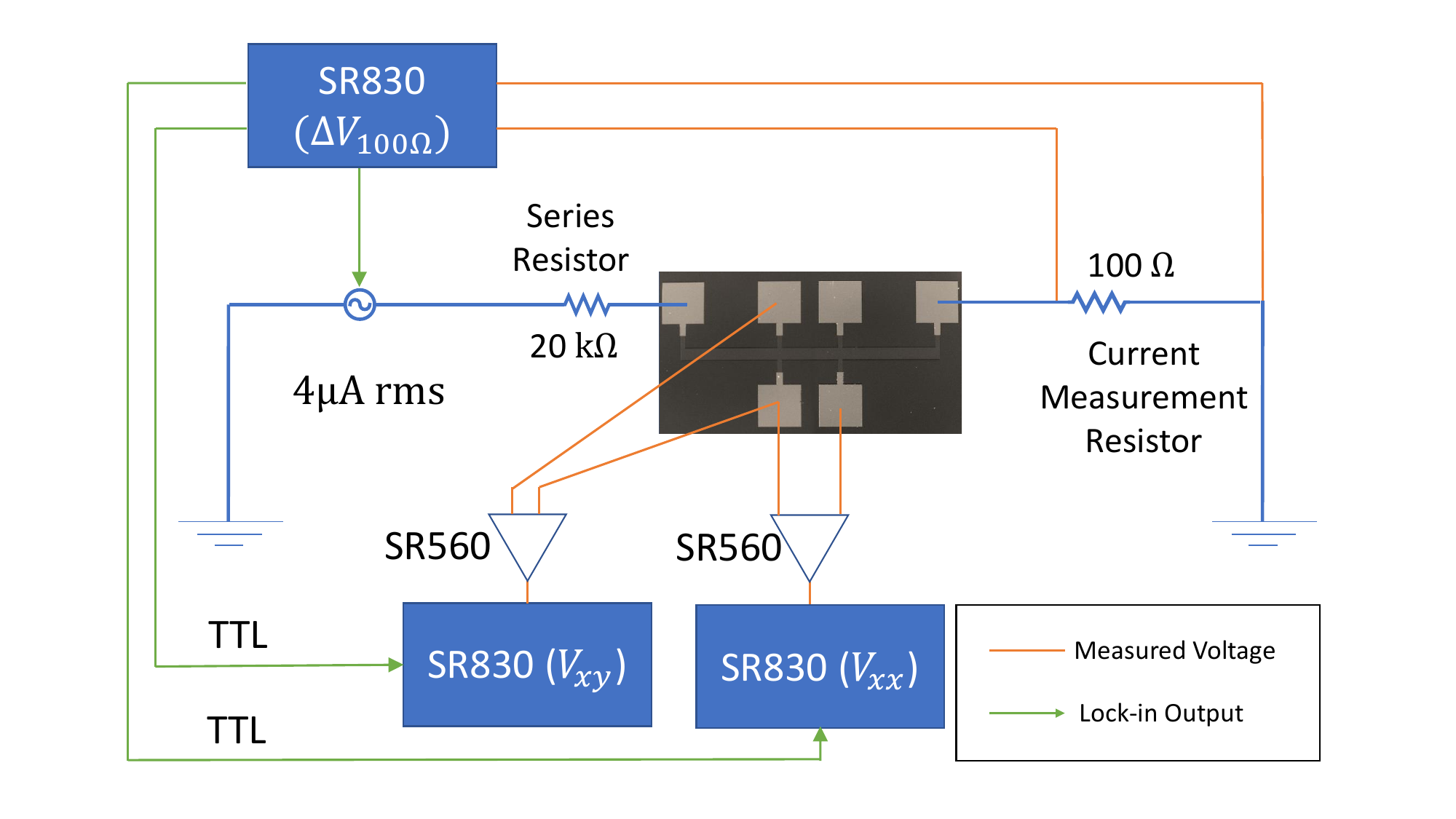}
    \caption{Full wiring diagram for the experimental setup for topological Hall resistivity measurements.}
    \label{fig:HallBarDiagram}
\end{figure}

\end{document}